# Phase diagrams on composition-spread $Fe_yTe_{1-x}Se_x$ films

Zefeng Lin[a,b], Sijia Tu[a,b], Juan Xu[a], Yujun Shi[a], Beiyi Zhu[a], Chao Dong[c], Jie Yuan[a,d], Xiaoli Dong[a,b,d], Qihong Chen[a,b], Yangmu Li[a,b], Kui Jin[a,b,d*], Zhongxian Zhao[a,b,d]

[a] *Beijing National Laboratory for Condensed Matter Physics, Institute of Physics, Chinese Academy of Sciences, Beijing 100190, China*
[b] *School of Physical Sciences, University of Chinese Academy of Sciences, Beijing 100049, China*
[c] *Institute of High Energy Physics, Chinese Academy of Sciences, Beijing 100049, China*
[d] *Songshan Lake Materials Laboratory, Dongguan 523808, China*

\* Corresponding author: Email: kuijin@iphy.ac.cn

**Abstract**

$Fe_yTe_{1-x}Se_x$, an archetypical iron-based high-temperature superconductor with a simple structure but rich physical properties, has attracted lots of attention because the two end compositions, Se content $x = 0$ and 1, exhibit antiferromagnetism and nematicity, respectively, making it an ideal candidate for studying their interactions with superconductivity. However, what is clearly lacking to date is a complete phase diagram of $Fe_yTe_{1-x}Se_x$ as functions of its chemical compositions since phase separation usually occurs from $x \sim 0.6$ to 0.9 in bulk crystals. Moreover, fine control of its composition is experimentally challenging because both Te and Se are volatile elements. Here we establish a complete phase diagram of $Fe_yTe_{1-x}Se_x$, achieved by high-throughput film synthesis and characterization techniques. An advanced combinatorial synthesis process enables us to fabricate an epitaxial composition-spread $Fe_yTe_{1-x}Se_x$ film encompassing the entire Se content $x$ from 0 to 1 on a single piece of $CaF_2$ substrate. The micro-region composition analysis and X-ray diffraction show a successful continuous tuning of chemical compositions and lattice parameters, respectively. The micro-scale pattern technique allows the mapping of electrical transport properties as a function of relative Se content with an unprecedented resolution of 0.0074. Combining with the spin patterns in literature, we build a detailed phase diagram that can unify the electronic and magnetic properties of $Fe_yTe_{1-x}Se_x$. Our composition-spread $Fe_yTe_{1-x}Se_x$ films, overcoming the challenges of phase separation and precise control of chemical compositions, provide an ideal platform for studying the relationship between superconductivity and magnetism.

## 1. Introduction

Iron-based superconductors have been a subject of intensive research since their discovery [1, 2], and yet the origin of superconductivity and non-trivial band topology is still under debate [3]. In this material family, iron chalcogenide $Fe_yTe_{1-x}Se_x$ has attracted special attention [4], partially due to its simple structure that only consists of the basic Fe(Te,Se) unit cell. $Fe_yTe_{1-x}Se_x$ has a relatively high bulk superconducting transition temperature ($T_c$) near 15 K at ambient pressure [5] and 40 K at high pressure [6]. Because of the similarities in the emergence of superconductivity from an antiferromagnetic (AFM) state between iron chalcogenides and cuprates, $Fe_yTe_{1-x}Se_x$ has been considered a good platform to explore the origin of high-$T_c$ superconductivity. Interestingly, recent studies reported features of topological surface superconductivity [7, 8] and Majorana bound states [9] in $Fe_yTe_{1-x}Se_x$, which made it a potential candidate for quantum computation. Owing to the significance of $Fe_yTe_{1-x}Se_x$ in the comprehension of high-$T_c$ mechanism and potential technological applications, it is important to understand the underlying physics of $Fe_yTe_{1-x}Se_x$.

One standing puzzle for $Fe_yTe_{1-x}Se_x$ is its electronic phase diagram. Due to the phase separation from $x \sim 0.6$ to 0.9 of the bulk crystals [10], a complete phase diagram of $Fe_yTe_{1-x}Se_x$ is yet to be constructed. In addition, a fine control of its chemical composition is difficult to achieve since volatile elements Te and Se [11, 12] lead to the extreme sensitivity of stoichiometry to the synthesis process. A subtle perturbation in sample growth would induce a considerable change in its chemical and physical properties and thus hamper elucidation of its superconductivity and band topology. To solve the problem of the metastable phase in bulk crystal growth, thin-film synthesis has been utilized. Pulsed laser deposition (PLD), in particular, is widely applied in obtaining high-quality $Fe_yTe_{1-x}Se_x$ films. Zhuang et al. and Imai et al. fabricated $Fe_yTe_{1-x}Se_x$ films that reside in the phase separation region[13, 14], suggestive of a new opportunity to obtain detailed information for a complete phase diagram. Currently, challenges still arise when investigating subjects such as the quantum phase transitions[15], for which fine control of the chemical composition is required. However, traditional film growth techniques that involve time-consuming trial-and-error synthesis protocols are no longer sufficient in coping with the research needs.

In the past two decades, the high-throughput combinatorial thin-film synthesis technology, which can achieve a continuous change of single chemical composition, has been well developed for the growth of complex materials [16-19]. Very recently, advanced combinatorial techniques of layer-by-layer growth have been successfully employed for the growth of combinatorial high-$T_c$ cuprate films. Wu et al. utilized a tilted Knudsen cell in an oxide molecular beam epitaxy system to fabricate composition-spread $La_{2-x}Sr_xCuO_4$ films and found evidence for competition between the superconducting state and a charge-cluster glass state near a quantum critical point [20]. Yuan et al. synthesized composition-spread $La_{2-x}Ce_xCuO_4$ films by combinatorial laser molecular beam epitaxy (CLMBE) with a programmable moving mask and discovered a numerical scaling relation between the scattering rate of the strange-metal state and the superconducting transition temperature [21, 22]. So far the application of this new technology to iron-based superconductors is still scarce.

In this paper, we apply composition-spread thin film synthesis and high-throughput characterization on the iron chalcogenide $Fe_yTe_{1-x}Se_x$. $Fe_yTe_{1-x}Se_x$ films ($x$ = 0 - 1) were synthesized on single-crystal $CaF_2$ substrates by CLMBE. Compositional analysis and X-ray diffraction reveal a continuous chemical doping and a smooth evolution of lattice parameters, respectively. Electrical transport measurements on microscales were performed to establish a complete electronic phase diagram of $Fe_yTe_{1-x}Se_x$ as a function of Se content. Our work demonstrates the advantages and feasibility of using composition-spread films in studying unconventional superconductors beyond cuprates.

## 2. Experimental methods

### 2.1. Sample preparation

The polycrystalline targets with nominal compositions of FeSe and FeTe were prepared with mixed high purity powders by solid-state reaction method. The mixing and grinding processes were carried out in the glovebox with $O_2$ and $H_2O$ levels both less than 1 ppm and then the mixed powders were compressed into pellets and sintered at 450°C for 24 hours in evacuated quartz tubes. For each target, grinding and sintering were repeated three times.

To fabricate the composition-spread $Fe_yTe_{1-x}Se_x$ film, the deposition parameters need to be cautiously controlled. We utilized the reflection high-energy electron diffraction (RHEED) to monitor the deposition rate and calibrated its value by synthesizing a single-component $Fe_yTe_{1-x}Se_x$ film. The 10×10 mm² (00$l$)-oriented single-crystal $CaF_2$ substrate was heated to 300 °C with the base pressure maintained at $10^{-9}$ Torr. The distance between the substrate and target was 50 mm. The fluence and repetition frequency of the KrF excimer laser ($\lambda$ = 248 nm) were ~ 5 J/cm² and 2 Hz, respectively. Periodic oscillations in the RHEED intensity were shown in Fig. 1a, suggestive of a layer-by-layer growth. According to the time interval of two adjacent peaks, 52 pulses correspond to the deposition of one unit cell (U.C.) $Fe_yTe_{1-x}Se_x$. A stripe-like RHEED pattern indicates the film surface is flat during growth (Fig. 1b).

Turning to the composition-spread $Fe_yTe_{1-x}Se_x$ film, the synthesis scheme is shown in Fig. 1c. The two targets were FeSe and FeTe, respectively. A mobile mask controlled the deposition area and thickness on the substrate via shielding partial plume from the ablated target. In the first half of the synthesis cycle, the FeSe target was rotated to 50 mm below the substrate with a forward movement of mask synchronized with the FeSe target ablation to deposit a wedge of FeSe. In the second half of the cycle, the FeTe target was rotated to 50 mm below the substrate with a backward movement of the mask to generate a FeTe wedge in the opposite direction. The combination of these two inverse wedges created a continuous distribution of $x$ as shown in Fig. 1d. In order to avoid a superlattice structure, the film thickness within one period should be constricted in a U.C. height. Therefore, the number of pulses within the half period was set to 52 according to the RHEED oscillation results mentioned above. As a result, we obtain an epitaxial composition-spread $Fe_yTe_{1-x}Se_x$ film ($x$ = 0 - 1) over 8.4 mm-length range on the substrate after 180 periods (i.e., total thickness ~ 100 nm).

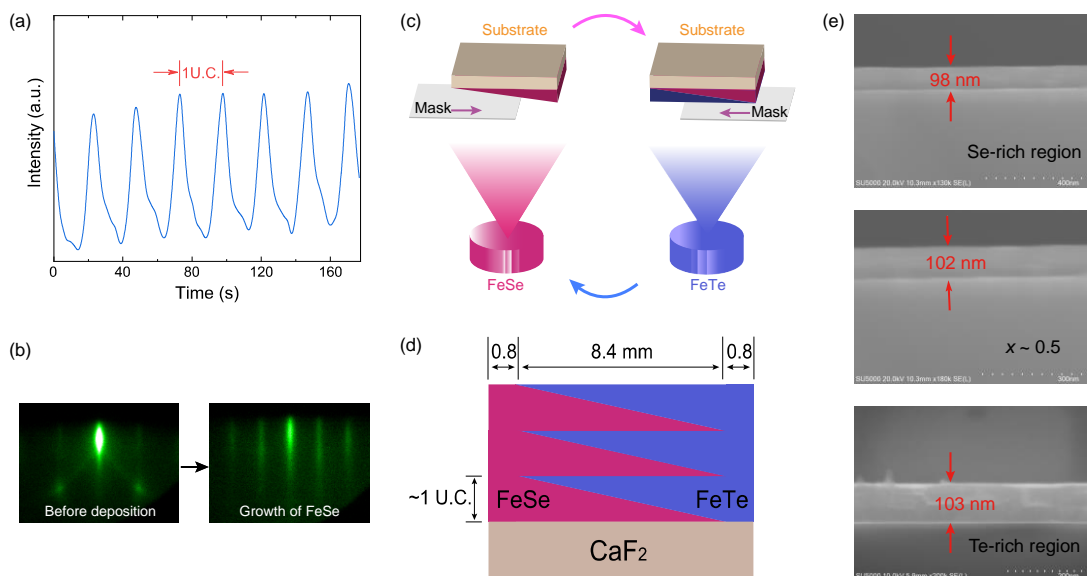

**Fig. 1.** (Color online) The growth of single-component and composition-spread $Fe_yTe_{1-x}Se_x$ films by the combinatorial laser molecular beam epitaxy. (a) RHEED intensity oscillations of epitaxial growth for FeSe films. The interval of two adjacent oscillation peaks corresponds to a unit-cell (U.C.) height. (b) RHEED patterns before deposition and during growth for FeSe films, indicative of the flat FeSe surfaces. (c) Schematic illustration of composition-spread film synthesis procedure within one period. Two targets with $x = 1$ and $x = 0$ are ablated alternately, during which a mask moves simultaneously to create two inverse wedges. (d) The schematic illustration of a stack for several inverse wedges. The desired thickness of a composition-spread film can be achieved by controlling the number of cycles. For convenience, $x$ spread is controlled over 8.4 mm-length range in the center on a 10 mm-length $CaF_2$ substrate. (e) SEM cross-sectional images at three different positions of the composition-spread $Fe_yTe_{1-x}Se_x$ film. The three panels from top to bottom represent the thicknesses of Se-rich, $x \sim 0.5$, and Te-rich regions, respectively.

*2.2. Sample characterization*

The film thickness was measured using a Hitachi SU5000 scanning electron microscopy (SEM). The energy-dispersive x-ray (EDX) spectrometer measurements were performed on several marked positions of the composition-spread film. We performed EDX analysis of $Fe_yTe_{1-x}Se_x$ films on $CaF_2$ substrates and observed large overlap between K edge of Ca and L edge of Te (Fig. S1a online). Similar issues have been reported on $CaF_2$ substrates and on $SrTiO_3$ substrates in previous studies[14, 23]. To avoid this problem, we also carried out EDX analysis on $Fe_yTe_{1-x}Se_x$ grown on the $LaAlO_3$ substrate (Fig. S1b online). The structure was characterized by a Rigaku SmartLab X-ray diffractometer (XRD) using the micro-region measurement component with a beam size of 0.2 mm. Electrical transport measurements were carried out using standard 4-probe method in the Physical Property Measurement System (PPMS).

## 3. Results and discussion

### 3.1. Film thickness, chemical concentration, and structure

Figure 1e shows the SEM cross-sectional images at different positions of the combinatorial $Fe_yTe_{1-x}Se_x$ film across the spread, all of which exhibit ~ 100 nm thickness, indicating a uniform deposition rate of FeSe and FeTe. The deposition rate is calculated to be 0.56 nm per cycle (the film thickness divided by number of cycles, 100 nm / 180). Note that 0.56 nm approximately corresponds to a U.C. thickness of $Fe_yTe_{1-x}Se_x$, consistent with Fig. 1d, which promotes homogeneous mixing of FeSe and FeTe and prevents superlattice formation.

The element ratio of composition-spread $Fe_yTe_{1-x}Se_x$ film (Fe:Te:Se) was mapped using EDX. The Se ($x$) and Fe ($y$) content vs. position across the film were plotted in Fig. 2a. The Se content varies linearly with position within the measurement uncertainty, spanning $0 \leq x \leq 1$. Here the uncertainty was estimated to be about $\pm 0.04$ from reproducible measurements. On the other hand, the Fe content ($y$) remains unchanged across the film. Similar to our results, the non-stoichiometry of $y$ ($y \sim 0.8$) in $Fe_yTe_{1-x}Se_x$ film has previously been reported using PLD by tuning the synthesis parameters [23, 24] or chemical composition of targets [25]. Nevertheless, the combinatorial film synthesis process allows us to fabricate a $Fe_yTe_{1-x}Se_x$ film with a uniform Fe content ($y$) and a position-dependent Se content ($x$) from 0 to 1 on a single piece of $CaF_2$ substrate.

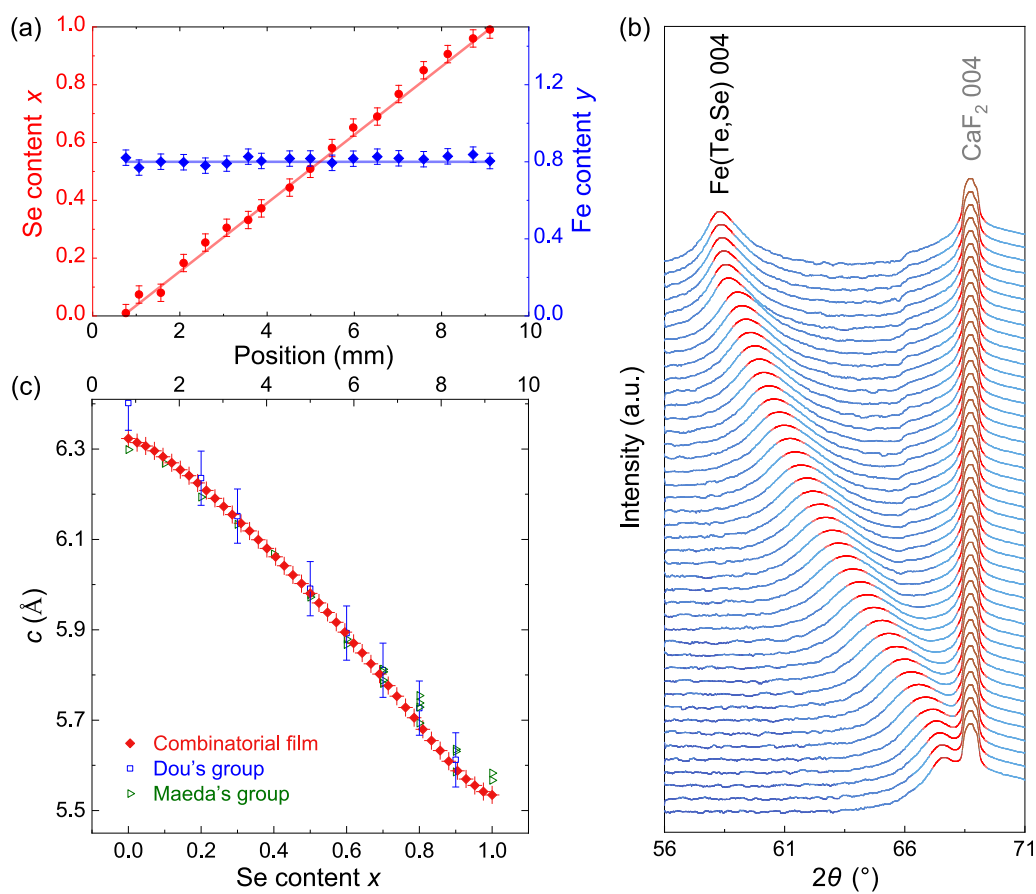

**Fig. 2.** (Color online) The compositional and structural characterization of the composition-spread $Fe_yTe_{1-x}Se_x$ films. (a) The Se doping level ($x$) (red circles) and Fe content ($y$) (blue diamonds) mapped across the composition-spread film by EDX spectroscopy. (b) The micro-region $\theta/2\theta$ XRD scan of the composition-spread film along the direction of composition spread. The $2\theta$ ranges from 56 to 71 degrees, showing the $Fe_yTe_{1-x}Se_x$ (004) and $CaF_2$ (004) peaks. The component interval $\Delta x$ between two adjacent curves is about 0.024. (c) The Se doping level ($x$) and position dependences of $c$-axis length (red diamonds) calculated from XRD data in (b). The blue squares and green triangles represent the $c$-axis lengths in single-component $Fe_yTe_{1-x}Se_x$ films from Ref. [13] and [14], respectively.

To gain the wealth of information in the composition spread at higher spatial resolution, the micro-region out-of-plane XRD with an $x$ resolution of 0.024 was performed on the composition-spread film. Figure 2b displays the $\theta/2\theta$ scan mapped at different positions with an interval of 0.2 mm along the direction of composition spread. The $Fe_yTe_{1-x}Se_x$ (004) diffraction peak moves to higher angles as $x$ increases while the $CaF_2$ (004) peak is fixed. There exist neither splitting nor broadening of the $Fe_yTe_{1-x}Se_x$ peaks even in Se-rich region $0.6 \leq x \leq 0.9$, where the phase separation usually occurs [10], indicating a good out-of-plane alignment at all positions. The suppression of the phase separation was previously observed in single component $Fe_yTe_{1-x}Se_x$ film [26]. Interestingly, we observed no phase separation in the composition-spread films on both $LaAlO_3$ and $SrTiO_3$ substrates, suggesting that the suppression of the phase separation is irrelevant to the lattice mismatch. As shown in Fig. 2c, the $c$-axis lattice constant monotonically decreases with $x$, consistent with that for single component $Fe_yTe_{1-x}Se_x$ films [13, 14]. For single chemical composition $Fe_yTe_{1-x}Se_x$ films, samples from different synthesis batches lead to large variations in $c$-axis length (see the open symbols in Fig. 2c). In contrast, for our $Fe_yTe_{1-x}Se_x$ film, the evolution of $c$-axis length is smooth, demonstrating a high accuracy in composition control.

*3.2. Electrical transport measurement*

As illustrated in Fig. 3a, the composition-spread $Fe_yTe_{1-x}Se_x$ film was patterned into microbridge arrays including 135 resistivity channels by photolithography, which achieved an $x$ resolution of 0.0074 over 60 μm-length microbridges. Note that this high resolution is the relative change for chemical compositions, since the absolute stoichiometry cannot be precisely measured using EDX. The temperature-dependent electrical resistivity ($\rho - T$) data from 100 to 2 K measured across the entire spread are shown in Fig. 3b, with superconducting transitions showing from $x \sim 0.1$ to $x \sim 1$. Three representative superconducting channels are depicted in Fig. 3c, which exhibit sharp superconducting transitions. Here, $T_c$ is defined as the intersection point by extrapolating the normal-state resistivity and the superconducting transition. As $x$ approaches 0 (i.e., the $Fe_yTe$ end), the superconducting transition vanishes while an obvious kink in $\rho - T$ occurs around 75 K in Fig. 3d, corresponding to the long-range antiferromagnetic (AFM) characteristic temperature ($T_N$) [10]. $T_N$ decreases gradually from $x \sim 0$ to $x \sim 0.1$, consistent with the results in single-crystal [27] and polycrystalline samples [10]. For FeSe, previous studies demonstrated that a kink signifying tetragonal-to-orthorhombic structural transition (nematic transition) appears in the $\rho - T$ of bulk sample near 90 K [28]. However, in film samples, this feature is absent both in the Fe$_y$Se end of our combinatorial film ($x \sim 1$ in Fig. 3e) and in the literature [26]. Instead, a peak is observed in the temperature dependence of the derivative of $\rho(T)$, $d\rho/dT$ versus $T$, as shown in Fig. 3f [26]. The temperature of this peak, defined as $T^*$ which is commonly accepted as the characteristic temperature for nematicity, decreases with reducing $x$.

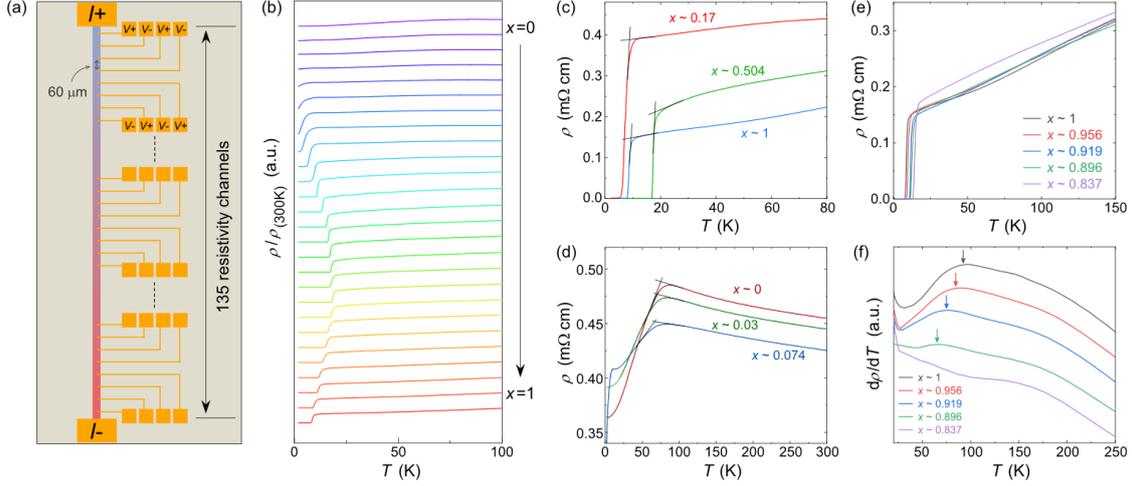

**Fig. 3.** (Color online) Micro-region electrical transport properties of the composition-spread Fe$_{0.8}$Te$_{1-x}$Se$_x$ film. (a) The schematic illustration of the patterned 135 bridges for transport measurements across the spread. The length of each microbridge is 60 μm, corresponding to an $x$ resolution of 0.0074. (b) Temperature dependence of normalized resistivity curves at different Se doping levels from 0 to 1. (c) Temperature-dependent resistivity of three representative superconducting Se doping level ($x \sim 0.17$, 0.504, and 1). $T_c$ is determined as the intersection point by extrapolation of normal-state resistivity and the superconducting transition. (d) Temperature-dependent resistivity of the extreme Te-rich doping levels ($x \sim 0$, 0.03, and 0.074). The long-range AFM characteristic temperature $T_N$ is determined as the position of the kink. (e) Temperature-dependent resistivity of the extreme Se-rich doping levels ($x \sim$ 0.837, 0.896, 0.919, 0.956, and 1). (f) Temperature differential of the normalized electrical resistivity extracted from (e).

*3.3. Phase diagrams and discussion*

Consequently, a two-dimensional phase diagram of resistivity was established. On the basis of the extracted $T_c$, $T_N$ and $T^*$ above, the corresponding superconducting, long-range AFM, and nematic phases are outlined in the phase diagram (Fig. 4a). In the Se-rich region, the resistivity in the normal state becomes smaller with raising $x$. In the extreme Te-rich region, the resistivity becomes larger and shows a small upturn before entering the superconducting state. The upturn behavior was reported to arise from weak charge carrier localization [27]. The superconducting region exhibits the typical dome shape in iron-based superconductors [13, 29-31]. The overall trend of the superconducting phase as a function of Se content looks similar to previous reports based on single-component films [13, 26] (see Fig. S2 online for a comparison among them), and yet some minor differences can be found in details of the $T_c$ evolution among these phase diagrams. For example, there is an obvious enhancement of $T_c$ for $0.6 < x < 0.8$ in Ref. [13], and a suppression of $T_c$ for $0.8 < x < 0.9$ in Ref. [26] which might be attributed to the nematic quantum criticality or in-plane strain [32]. In our experiment, $T_c$ varies smoothly as a function of $x$ and shows the highest value at $x \sim 0.7$, different from the maximum value observed at $x \sim 0.8$ in the previous study [26].

There are some possible origins for the observed minor differences of the reported phase diagrams. First of all, most published work usually assumed that the compositions of $Fe_yTe_{1-x}Se_x$ films were equivalent to that of targets, but the different synthesis parameters in PLD may affect the chemical ratio of Se to Te, which has a considerable impact on the evolution of superconductivity. For instance, by changing the growth temperature, Seo et al. fabricated a series of $Fe_yTe_{1-x}Se_x$ films with Se composition ranging from $x = 0.64$ to $0.72$ using the same target with $x \sim 0.45$ [23]. Secondly, Fe content can have a considerable effect on the phase diagram. To clarify the effect of Fe content on films, we tuned the Fe content of the composition-spread $Fe_yTe_{1-x}Se_x$ film from $y \sim 0.8$ ($Fe_{0.8}Te_{1-x}Se_x$) to $y \sim 0.74$ ($Fe_{0.74}Te_{1-x}Se_x$) by changing the laser fluence to $\sim 2.8$ J/cm$^2$. Similar to the $Fe_{0.8}Te_{1-x}Se_x$ film, the Fe content $y$ across the $Fe_{0.74}Te_{1-x}Se_x$ film maintained excellent uniformity, as shown in Fig. S3 (online). By controlling $x$ and $y$, we establish a three-dimensional phase diagram in Fig. 4b. Compared to the $Fe_{0.8}Te_{1-x}Se_x$ film, the slight decrease of Fe content $y$ in the $Fe_{0.74}Te_{1-x}Se_x$ film ($\Delta y \sim 0.06$), leads to the $T_c$ suppression of $\sim 2$ K of the whole dome i.e., $\Delta T_c \sim 2$ K. So, the suppression of $T_c(y)$ is estimated to be about 3.3 K/0.1(Fe), similar to $T_c(x)$, which was estimated to be about 3 K/0.1(Se). Additionally, the optimal $x$ with the highest $T_c$ shifts from $\sim 0.6$ to $\sim 0.7$ as $y$ changes from 0.74 to 0.8. Recently, Zhang et al. found a significant Fe content deviation between the nominal targets and single-component $Fe_yTe_{1-x}Se_x$ films [33], which was included in Fig. 4b. In our composition-spread film, we ensure a constant Fe content with chemical characterization at different film positions, keeping Se content as the single variable in a two-dimensional phase diagram. By changing the Fe content, this technique allows us to efficiently construct a specific two-dimensional path on the three-dimensional phase diagram.

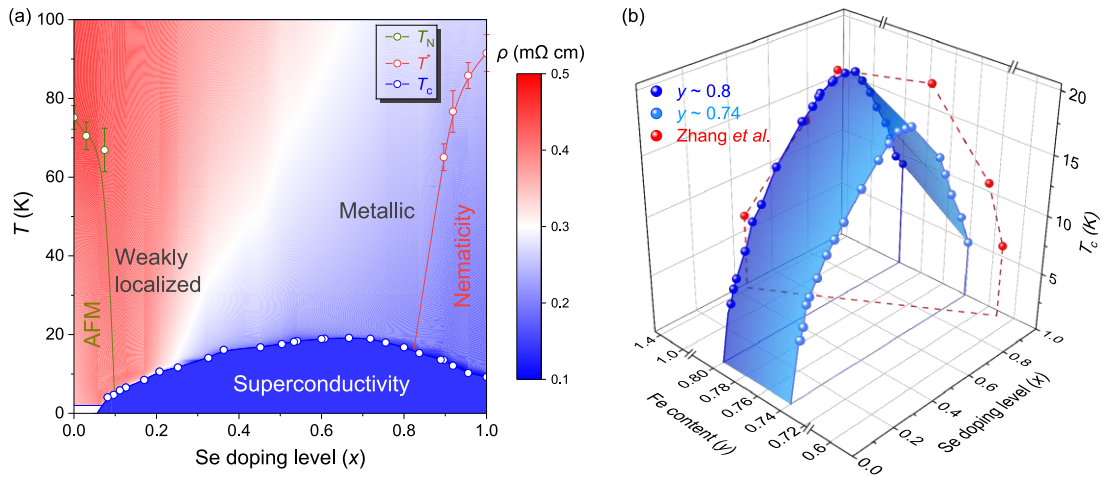

**Fig. 4.** (Color online) (a) The two-dimensional phase diagram of resistivity as functions of temperature and Se doping level ($x$). The open blue circles refer to the values of $T_c$ for the composition-spread $Fe_{0.8}Te_{1-x}Se_x$ film. The blue line connecting the values of $T_c$ shows the superconducting regime. The open olive circles represent the values of $T_N$. The long-range AFM phase is marked by the olive line. The open red circles represent the values of $T^*$. The nematic regime is marked by the red line. (b) The three-dimensional $T_c$ phase diagram as functions of Se doping level ($x$) and Fe content ($y$). The blue and cyan lines outline $T_c$ domes of the composition-spread $Fe_{0.8}Te_{1-x}Se_x$ and $Fe_{0.74}Te_{1-x}Se_x$ films, respectively. The red circles refer to $T_c$ values of single-component $Fe_yTe_{1-x}Se_x$ films from Ref. [33], and the superconducting regime is surrounded by the red dashed line.

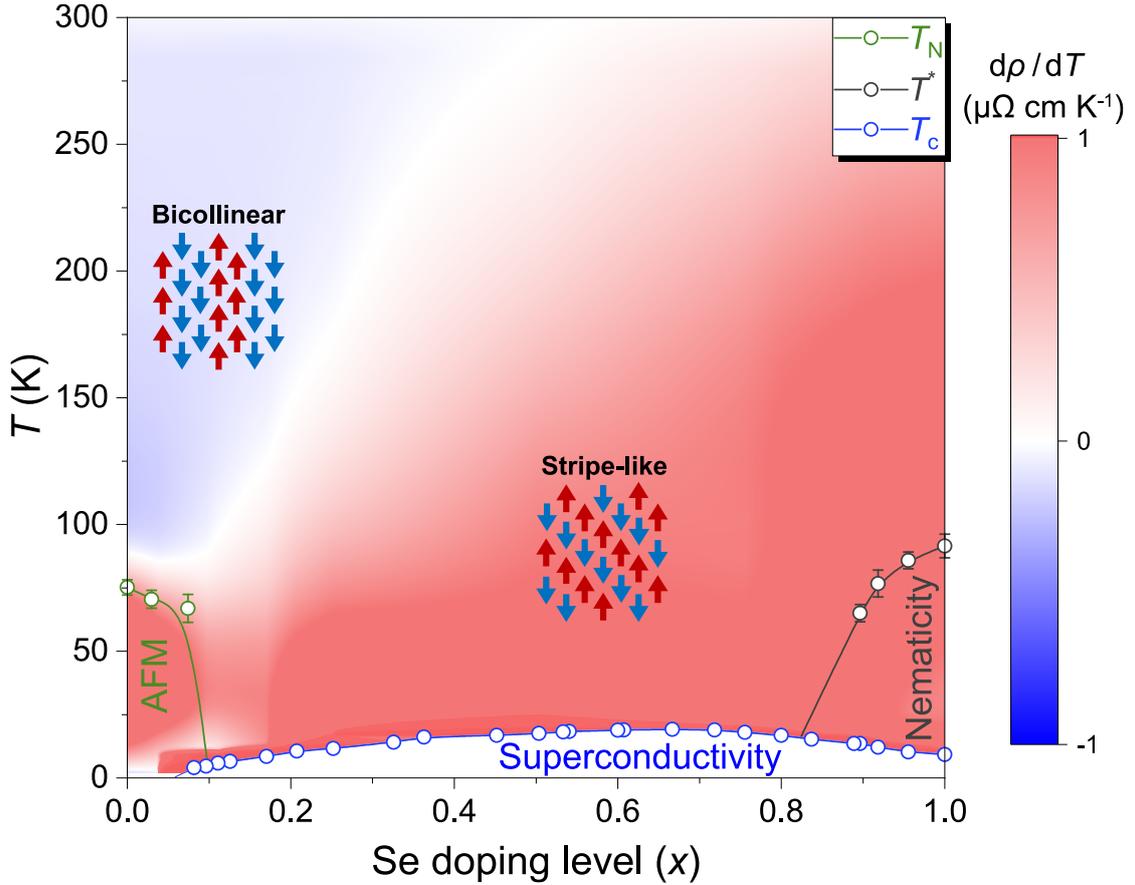

**Fig. 5.** (Color online) The two-dimensional phase diagram of differential resistivity ranging from 2 K to 300 K as a function of Se doping level ($x$). The short-range spin patterns in $Fe_yTe_{1-x}Se_x$ from Ref. [34] are added in the diagram.

In multi-band iron-based superconductors, there exists a variety of magnetic short-range spin patterns, which are closely related to the emergence of superconductivity. Neutron scattering observed that a bicollinear spin pattern usually coexists with an insulating behavior and a stripe-like spin pattern coexists with superconductivity in $Fe_yTe_{1-x}Se_x$ [7]. Angle-resolved photoemission spectroscopy (ARPES) studies show that, by forming different spin patterns, the self-energy and band coherence of $Fe_yTe_{1-x}Se_x$ can be altered accordingly [34, 35]. With a bicollinear spin pattern at intermediate $x$ and high temperature, the electronic bands resemble those of $Fe_yTe_{1-x}Se_x$ with small

Se content, which can suppress superconductivity [36]. In addition, photoinduced similar changes in the electronic band structure were reported, indicating an intrinsic relationship between magnetic spin patterns, band structure, and superconductivity [37]. In $Fe_yTe_{1-x}Se_x$, the change of spin patterns and band coherence can be characterized by a weak hump in resistivity [27], as different bands contribute differently to the orbital-selective Mott phase [38].

A color mapping of $d\rho/dT$ versus $T$ and $x$ for the composition-spread $Fe_{0.8}Te_{1-x}Se_x$ film was drawn in Fig. 5. Here we overlay the spin patterns obtained by neutron scattering studies in Fig. 5 for comparison [34]. Surprisingly, our $d\rho/dT$ results match the phase space for different spin patterns rather well, even though neutron experiments were performed on bulk crystals and our work on a composition-spread thin film. The long-range AFM phase at $x \sim 0$ is clearly observed in $d\rho/dT$. For the broader phase space, $d\rho/dT$ can be divided into two regions (negative and positive). The white boundary in between implies a change from an insulating to metallic behavior. For $x \sim 0$, or at higher temperatures when $x$ increases, $d\rho/dT < 0$ (insulating) and a bicollinear spin pattern exists. As $x$ increases and temperature falls further, $d\rho/dT > 0$ (metallic) and a stripe-like spin pattern emerges. This coincidence is likely associated to the itinerant and local origin of the magnetism and demonstrates the advantages of using the high-throughput composition-spread thin films, for which a complete phase diagram can be established using one sample. It has been argued that selenium substitution tunes the magnetic correlations, and the dynamic stripe-like spin pattern is necessary for superconductivity [34]. However, in our work, the temperature of transition from bicollinear to stripe-like spin pattern (the white boundary) increases monotonically with raising Se content $x$, while the evolution of $T_c$ is non-monotonic with $x$. The superconductivity in $Fe_yTe_{1-x}Se_x$ is not simply controlled by mobility and carrier density of its normal state [39-41], which requires further work to clarify.

## 4. Conclusion

To summarize, we report a fine chemical concentration study of the electronic properties of $Fe_yTe_{1-x}Se_x$ achieved by advanced high-throughput film synthesis and characterization techniques. A unique combinatorial synthesis process allows the epitaxial growth of a composition-spread $Fe_yTe_{1-x}Se_x$ film with a uniform Fe content ($y$) and a position-dependent Se content ($x$) from 0 to 1 on a single piece of $CaF_2$ substrate. Micro-region structural characterization of our composition-spread $Fe_yTe_{1-x}Se_x$ film shows the absence of phase separation, which occurs from $x \sim 0.6$ to 0.9 in bulk crystals. By comparing two combinatorial films with different $y$ values, we demonstrate the important role of Fe content on its superconductivity, namely, different Fe contents result in an overall shift of the superconducting dome in the Se content axis, which might be correlated with the discrepancies in the previously reported phase diagrams. Based on micro-region electrical transport measurements, we have established a two-dimensional phase diagram, which provides a $d\rho/dT$ mapping associated with the evolution of spin patterns. Further studies are required to clarify the emergence of superconductivity in $Fe_yTe_{1-x}Se_x$, and the interactions between superconductivity and electronic states such as antiferromagnetism and nematicity. As the first demonstration of high-throughput methodology on iron-based superconductors, our work opens a new direction for these tasks.

**Conflict of interest**

The authors declare that they have no conflict of interest.

**Acknowledgement**

This work is supported by the National Key R&D Program of China (Grants No. 2021YFA0718700, No. 2017YFA0302902, No. 2017YFA0303003, and No. 2018YFB0704102), the National Natural Science Foundation of China (Grants No. 11834016, No. 11961141008, No. 11927808, and No. 12174428), the Strategic Priority Research Program (B) of Chinese Academy of Sciences (Grants No. XDB25000000, and No. XDB33000000), the Beijing Natural Science Foundation (Grants No. Z190008), CAS Interdisciplinary Innovation Team, Key-Area Research and Development Program of Guangdong Province (Grant No. 2020B0101340002), and the Center for Materials Genome.

**Author contributions**

Zhongxian Zhao, Kui Jin and Xiaoli Dong conceived the project. Zefeng Lin and Yujun Shi prepared the samples guided by Kui Jin and Jie Yuan. Zefeng Lin and Sijia Tu performed structural and transport measurements. Juan Xu performed the SEM measurements. Kui Jin, Yangmu Li, Qihong Chen, Jie Yuan, Chao Dong, Beiyi Zhu, Zefeng Lin and Sijia Tu analyzed the data and wrote the manuscript with contributions from all the authors.

Supplementary Information

# Phase diagrams on composition-spread Fe$_y$Te$_{1-x}$Se$_x$ films

Zefeng Lin[a,b], Sijia Tu[a,b], Juan Xu[a], Yujun Shi[a], Beiyi Zhu[a], Chao Dong[c], Jie Yuan[a,d], Xiaoli Dong[a,b,d], Qihong Chen[a,b], Yangmu Li[a,b], Kui Jin[a,b,d*], Zhongxian Zhao[a,b,d]

[a] *Beijing National Laboratory for Condensed Matter Physics, Institute of Physics, Chinese Academy of Sciences, Beijing 100190, China*
[b] *School of Physical Sciences, University of Chinese Academy of Sciences, Beijing 100049, China*
[c] *Institute of High Energy Physics, Chinese Academy of Sciences, Beijing 100049, China*
[d] *Songshan Lake Materials Laboratory, Dongguan, Guangdong 523808, China*

\* Corresponding author: Email: kuijin@iphy.ac.cn

## 1. EDX results

An overlap between the energy peaks of K edge of Ca and L edge of Te as shown in Fig. S1(a). It affects the composition analysis for Fe$_y$Te$_{1-x}$Se$_x$ films on CaF$_2$ substrates using the energy-dispersive x-ray (EDX) spectroscopy, as reported in references [1] and [2]. Similar problem exists in Fe$_y$Te$_{1-x}$Se$_x$ films on SrTiO$_3$ substrates. Fortunately, there is no energy overlap in Fe$_y$Te$_{1-x}$Se$_x$ films on LaAlO$_3$ substrates in Fig. S1(b). In our work, a composition-spread Fe$_y$Te$_{1-x}$Se$_x$ film grown on a LaAlO$_3$ substrate instead of the composition-spread Fe$_y$Te$_{1-x}$Se$_x$ film on the CaF$_2$ substrate was utilized for composition analysis.

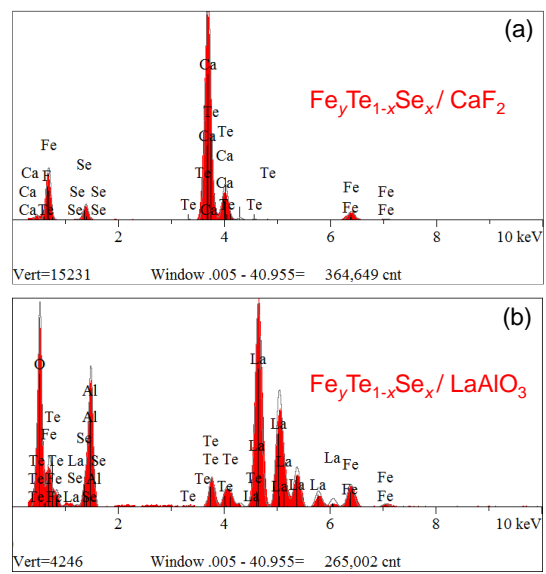

Fig. S1. The energy-dispersive x-ray (EDX) spectroscopy results of Fe$_y$Te$_{1-x}$Se$_x$ films on CaF$_2$ (a) and Fe$_y$Te$_{1-x}$Se$_x$ films on LaAlO$_3$ (b).

## 2. Comparison between $T_c$ phase diagrams

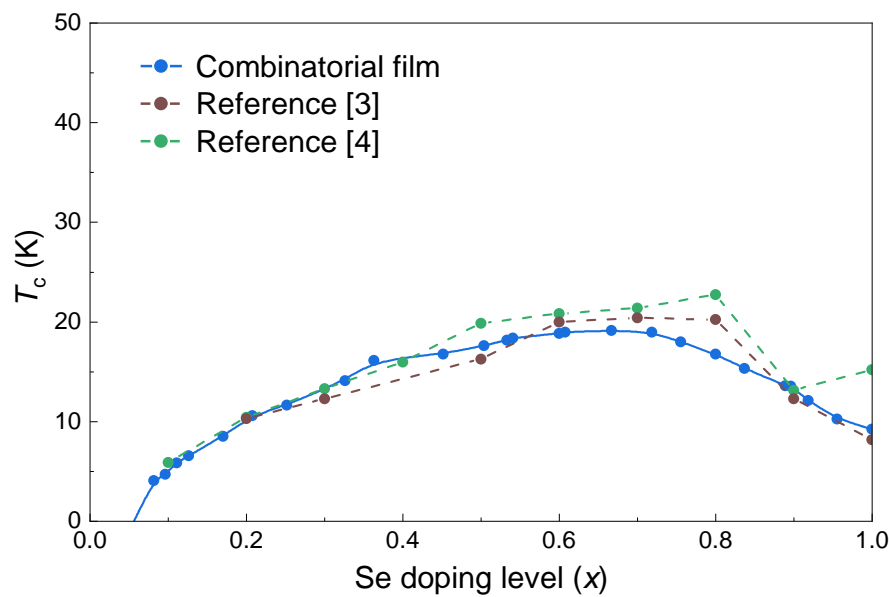

Fig. S2. Three $T_c$ phase diagrams as a function of Se concentration ($x$), including the values of $T_c$ for the composition-spread $Fe_{0.8}Te_{1-x}Se_x$ film (blue circles), the single-component films in references [3] (brown circles) and [4] (green circles).

## 3. Composition and electrical transport results of the composition-spread $Fe_{0.74}Te_{1-x}Se_x$ film

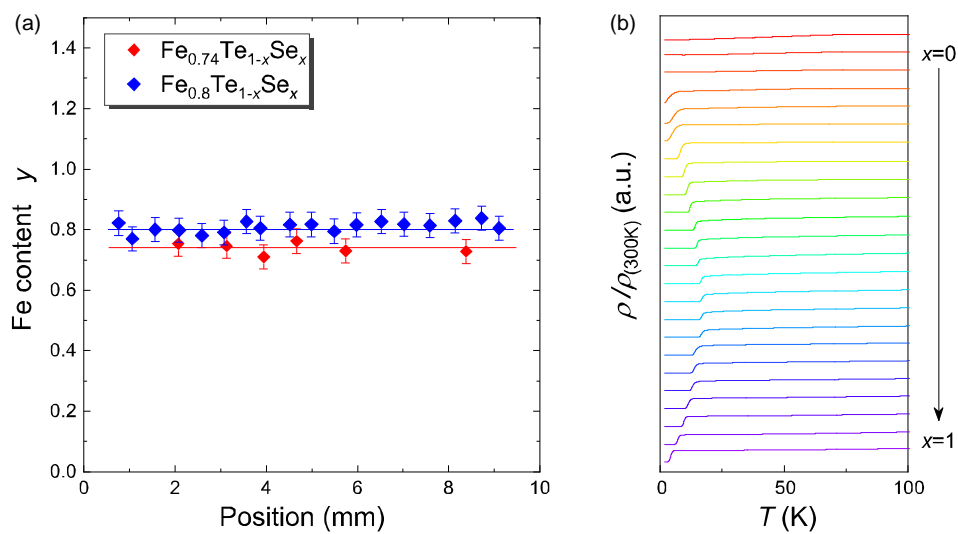

Fig. S3. Fe content $y$ and electrical transport properties of the composition-spread $Fe_{0.74}Te_{1-x}Se_x$ film. (a) The Fe content $y$ across the composition-spread $Fe_{0.74}Te_{1-x}Se_x$ film (red diamonds) and $Fe_{0.8}Te_{1-x}Se_x$ film (blue diamonds). (b) Temperature dependence of normalization resistivity at different Se concentrations of the composition-spread $Fe_{0.74}Te_{1-x}Se_x$ film by micro-region electrical transport measurement. The Se doping level $x$ varies from 0 to 1.